\documentclass{emulateapj}
\newcommand{\etal}{et al.\ }
\newcommand{\etalb}{et al.}
\newcommand{\be}{\begin{equation}}
\newcommand{\ba}{\begin{eqnarray}}
\newcommand{\ee}{\end{equation}}
\newcommand{\ea}{\end{eqnarray}}

\begin{document}
\title{Three-Body Kick to a Bright Quasar out of Its Galaxy During a Merger}

\author{Loren Hoffman\altaffilmark{1} and Abraham 
Loeb\altaffilmark{2}}

\email{lhoffman@cfa.harvard.edu; aloeb@cfa.harvard.edu}

\altaffiltext{1}{Physics Department, Harvard University, 17 Oxford St.,
Cambridge, MA 02138}

\altaffiltext{2}{Astronomy Department, Harvard University, 60 
Garden Street, Cambridge, MA 02138}


\begin{abstract}
The quasar HE0450-2958 was recently discovered \citep{M05} 
to reside $\sim 7$ kpc away from a galaxy that was likely disturbed by a
recent merger. The lack of a massive spheroid of stars around the quasar
raised the unlikely suggestion that it may have formed in a dark
galaxy. Here we explain this discovery as a natural consequence of a
dynamical kick imparted to the quasar as it interacted with a binary black
hole system during a galaxy merger event. The typical binary stalling radius
provides a kick of order the escape velocity of the stellar spheroid, 
bringing the quasar out to around the observed radius before it turns 
around. This is consistent with the observed low relative velocity between 
the quasar and the merger-remnant galaxy.  The gas carried with the black 
hole throughout the three-body interaction fuels the quasar for the 
duration of its journey, $\sim 2\times 10^7$ years.  Gravitational 
radiation recoil could not have produced the required kick.
\end{abstract}

\section{Introduction}
Observations indicate that every massive galaxy contains a supermassive
black hole in its nucleus, which shines as a quasar during episodes of
rapid mass accretion.  The vigorous gas inflows needed to fuel the most
luminous quasars are thought to be triggered by galaxy mergers, in which
tidal interactions channel large amounts of gas into the central $\sim 100$
pc \citep{BSV87,H89}.  The maximum luminosities of quasars are
strongly correlated with the properties of their host
galaxies \citep{S03,FL04}, in accordance with the tight
correlations observed between black hole mass and stellar bulge
mass \citep{M98,M03,P05} and velocity
dispersion \citep{FM00,G00,T02} in quiescent galaxies.

Given these correlations, the recent discovery \citep{M05} of a bright 
($M_{V}=-25.8$) quasar at redshift $z=0.285$ apparently {\it not} surrounded 
by a massive host galaxy is surprising indeed.  For emission at the Eddington 
luminosity, the inferred black hole mass is $\sim 4\times10^{8}M_{\odot}$.  
There is a neighboring, luminous starburst galaxy at a separation of 
$\sim 1.7''$ corresponding to 7.2 kpc for a flat universe with a present-day 
Hubble parameter $H_{0}=71~{\rm km~s^{-1}~Mpc^{-1}}$, and a matter density 
parameter $\Omega_{m}=0.27$. This companion galaxy has a disturbed morphology 
and is heavily obscured by dust, typical of the Ultra-Luminous Infrared 
Galaxies (ULIRGs) systematically associated with near equal-mass mergers 
\citep{S88}.  Its visual magnitude is $M_{V}=-23$.  Close to the
quasar lies a bright blob about 2 kpc across, consisting of gas ionized
by the quasar radiation.  The spectrum of the blob shows no stellar
component and no dust extinction; hence the host galaxy cannot be hidden by
dust.  The typical host for an $M_{V}=-25.8$ quasar is an elliptical galaxy 
with $M_{V}\approx -23.2$, about the magnitude of the neighboring ULIRG.  
However after carefully deconvolving a new HST image of the system, Magain 
\etal find no sign of any such galaxy surrounding this quasar. Fixing the
half-light radius at $R_{e}=10$ kpc (a typical size for the host of an 
$M_{V}\sim -26$ quasar), they put a conservative upper limit of
$M_{V}=-21.2$, $3.7\sigma$ fainter than expected, on the magnitude of any 
host.  Alternatively, adopting $M_{V}=-23$ for the host magnitude, they set 
an upper limit of $\sim 100$ pc on $R_{e}$, much too small for a normal 
galaxy of this magnitude.

We suggest that the quasar observed in HE0450-2958 may be a black hole that
was ejected from the nucleus of the companion galaxy during the major
merger causing its disturbed appearance.  The gas now fueling the quasar
was dragged with it from the galactic center when it was kicked out.  
If one of the merging galaxies harbored a binary black hole, then when the 
black hole from the other galaxy entered the system a complex 3-body 
interaction would ensue, shortly ending with the ejection of the lightest 
black hole at a velocity of order the binary orbital velocity \citep{H75}.  
This scenario implies that one of the merging galaxies was relatively 
gas-poor (perhaps a giant elliptical), to allow survival of the binary 
against gasdynamical friction for a Hubble time before the merger.  On the 
other hand at least one of the galaxies must have been gas-rich, in order 
to fuel the quasar.  The two galaxies must have been close in mass,
to explain the ULIRG appearance of the merger remnant and the high inferred
black hole mass of the quasar, since it is the lightest body that gets 
ejected.

The Eddington accretion rate for a black hole of mass $M=4 \times 10^{8}$
$M_{\odot}$ is $\dot{M}_{Edd}=(4\pi GM m_{p})/(\epsilon \sigma_{T}c)= 8.9
\epsilon_{-1}^{-1} M_{\odot}~{\rm yr}^{-1}$, where $m_{p}$ is the proton
mass, $\sigma_{T}$ the Thomson cross-section, and
$\epsilon=0.1\epsilon_{-1}$ is the radiative efficiency. It takes $\sim 2
\times 10^{7}$ yr for the black hole to reach a turnaround radius of 7.2
kpc, so in order to remain shining with $\epsilon_{-1}\sim 1$ at the
observed distance the black hole must drag at least $\sim 36\%$ of its own
mass with it in gas when it is ejected.  The distances
of closest approach in our three-body interactions are often larger than
the radius containing this mass in an $\alpha$--disk \citep{ST83}, and so the
inner accretion disk can follow the black hole throughout the interaction
and get ejected with it.

\section{Black hole binaries and ejections}
When two galaxies of comparable mass merge, their black holes sink to the
center by dynamical friction and form a gravitationally bound binary.  Once
the binary separation reaches the "hardening radius", $r_{hard}\approx G
[\min(M_{1},M_{2})]/4 \sigma^{2}\sim$(1--5) pc for massive galaxies (where
$\sigma$ is the 1D stellar velocity dispersion and $M_{1,2}$ are the black
hole masses), its further reduction is dominated by close encounters with
stars that pass within the binary orbit.  These stars undergo strong 3-body
interactions with the binary and escape its vicinity with speeds comparable
to the binary orbital speed.  Only stars on nearly radial orbits can
contribute to the hardening of the binary in this stage.  In the
low-density nuclei of massive elliptical galaxies, the total mass in stars
on these so-called "loss cone" orbits is small compared to the mass of the
binary.  Since the two-body relaxation time is long compared to the Hubble
time, stars cannot diffuse back into the loss cone as fast as they are kicked 
out.  If all stars on loss cone
orbits run out, the decay of the binary ceases.  Since the binary ejects of
order its own mass in stars per $e$-folding in radius, one naively expects
binaries in the cores of massive elliptical galaxies to stall at
separations of order 1 pc.  If the separation reaches $\sim$0.01-0.1pc,
then the binary would quickly coalesce through the emission of
gravitational radiation.  The question of whether and how it crosses the
gap from $\sim$ 1 pc to this smaller scale is known as the "final parsec
problem" \citep{MM04}.

Black hole binaries may coalesce in some cases but not others. For example,
the mass of any close binary companion to the nuclear black hole of the
Milky Way is limited to $M_{2}\la 10^{4}M_{\odot}$ by the Keplerian orbits
of stars near the galactic center \citep{MM04}.  The so-called X-shaped radio
galaxies have been explained as jets bent by sudden changes in the spin of
a black hole upon coalescence with a binary companion \citep{MM04}.  Gas drag
may lead to coalescence in gas-rich mergers \citep{MM04}; however the
gas-poor cores of massive ellipticals could allow binaries to stall.  There
are some tentative observational hints that sub-parsec scale binaries exist,
including periodic variations and double-peaked broad line profiles in
active galaxies, and central minima in the surface brightness profiles of
some early type galaxies which are possibly associated with the clearing of
the loss cone around stalled binaries \citep{K03}.

When an unequal mass black hole binary coalesces by gravitational
radiation, the waves carry off a net linear momentum and the merged black
hole recoils in the opposite direction \citep{FHH04}.  However 300 km
s$^{-1}$ is a reasonable upper limit to the velocity achievable by this
mechanism \citep{M04,BQW05}.  A mass ejected radially at 300 km
s$^{-1}$ from the center of the average stellar spheroid hosting a $4
\times 10^{8} M_{\odot}$ black hole ($M_{vir}=1.7 \times 10^{11}M_{\odot}$,
$R_{e}=3.6$ kpc) would only reach a distance of $\sim 150$ pc before
turning around, far smaller than the separation of the quasar in
HE0450-2958 from its companion galaxy.

If a binary stalls and its host galaxy merges with another galaxy, a third
black hole may spiral in and undergo strong three-body interactions with
the pair.  The merger event will drive mass through the binary at a
significant rate, partially refilling the loss cone, but the final stages
of the merger transpire so rapidly that the intruder is likely to arrive in
the vicinity of the binary before the binary has a chance to
coalesce \citep{MM01}.  The three-body interaction will end in 
the hardening of the binary by an amount comparable to its initial binding 
energy, and the ejection of the lightest body at a speed comparable to the 
binary's orbital speed \citep{H75}.  The binary's orbital speed at the 
stalling radius is a few times $\sigma$, i.e. of order the escape 
velocity $v_{esc}$ from the stellar bulge (e.g. for a Hernquist 
profile \citep{H90} $v_{esc}=2 \sqrt{2} v_{circ}(r=a)$ with $\sigma \sim 
v_{circ}$).  Thus, qualitatively we expect three-body kicks to yield 
ejection velocities of order the bulge escape velocity, resulting in many 
near-ejections.  This is consistent with the large distance and low velocity
of the HE0450-2958 quasar relative to the ULIRG.  Figure 1 shows an example 
of such a near-escape trajectory.

\begin{figure} 
\plotone{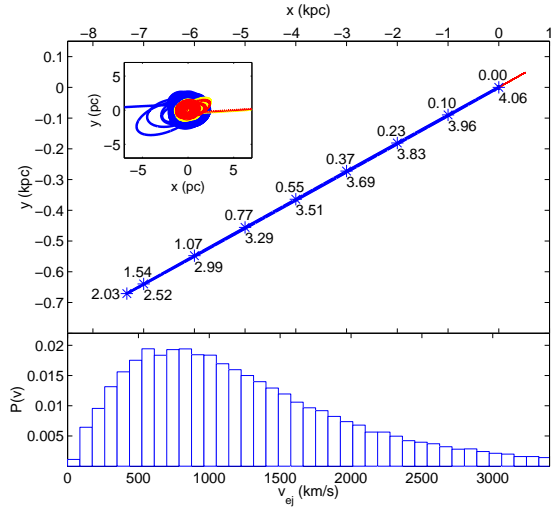} 
\caption{Black hole trajectories following a 3-body ejection.  The blue
    (thick) line shows the path of the intruder and the red and yellow
    (thin) lines that of the binary members; each is plotted only until its
    first return to the galactic center.  Time labels in units of $10^{7}$
    yrs are marked off at various points along the trajectory, going out
    and returning.  In the upper left inset we zoom in on the 
    innermost $\pm 7$ pc.  Here the trajectories are plotted only up to the 
    point of ejection (with return trajectories left out).  One can clearly 
    see the binary and intruder flying off in opposite directions at the 
    end of the interaction.  The lower panel shows the distribution
    of ejection velocities over all $\sim$10$^{5}$ runs.}
\end{figure}

\section{Numerical calculations}
To make our dynamical arguments more quantitative, we ran a series of
83,250 3-body scattering experiments using Sverre Aarseth's
KS-regularized \citep{KS65} Chain code \citep{A03} to compute the
black hole trajectories.  In each case the intruder's mass was set to
$M_3=4 \times 10^{8}M_{\odot}$ and the masses of the initial binary members
were chosen uniformly in their logarithm between 1.0 and 3.5 times the mass
of the intruder, a generic range for major mergers.  From here on we will
refer to the initial black hole binary as the inner binary and the
``binary'' formed by the intruder and the inner binary's center of mass as
the outer binary.  The semi-major axis $a_{bin}$ of the inner binary was 
always initialized to $r_{hard}/4$, and that of the outer binary was set 
using the stability criterion of Mardling and Aarseth \citep{MA99},
$({R_{p}^{out}}/{a_{in}}) \approx 2.8 [(1+q_{out})
{(1+e_{out})}/{\sqrt{1-e_{out}}}]^{\frac{2}{5}}$, which reliably estimates
the most distant intruder orbit for which strong 3-body interactions can
begin.  Here $R_{p}^{out}$ is the pericentre distance of the outer binary,
$a_{in}$ is the semimajor axis of the inner binary,
$q_{out}=M_{3}/(M_{1}+M_{2})$ is the mass ratio of the outer binary, and
$e_{out}$ is the outer binary's eccentricity.  The initial eccentricity of
the inner (outer) binary was chosen uniformly between 0.0 and and 0.2
(0.3), where this range of low eccentricities naturally results from
dynamical friction.  Finally, the three Euler angles of the intruder's
orbital plane were chosen randomly relative to the reference plane of the
binary orbit, as was the phase of the initial periastron of the binary.
Both orbits were always started at pericentre; since many orbital periods
elapse before unstable interactions begin, the relative phase is
effectively randomized in any case.

The Chain code allows the user to add conservative or velocity-dependent
external perturbations, and so we added a dynamical friction force given by
Chandrasekhar's formula \citep{C43} to the motion of each body.  The density
and Coulomb logarithm $\ln \Lambda$ were chosen to give a dynamical
friction timescale $\sim 10^{6}$ yr.  Each run was terminated when (a) any
one of the bodies was expelled to a distance of 1 kpc from the center of
mass; (b) any two bodies came within $30 r_s$ (where $r_s=2GM/c^2$ is the
Schwarzschild radius) of each other or the gravitational radiation
timescale dropped below $10^{5}$ yr (roughly the dynamical time) for any
pair, or (c) the total time elapsed reached $6 \times 10^{6}$ yr.  We
selected out just those runs where the intruder never came within 5000
Schwarzschild radii ($\sim$0.2 pc) of either binary member, reflecting the
scale below which an $\alpha$--disk \citep{ST83} with sufficient mass to
fuel the quasar might be disrupted.  This is also about the size inferred 
for the broad-line region of the HE0450-2958 quasar \citep{GH05, MER05}. Of 
the runs not terminated by criterion (b) or (c), the closest approach of the
intruder was $>5\times 10^3r_{s}$ in 36\% and $>10^4r_{s}$ in 22\% of the
runs.  To each successful run we then assigned a galaxy based on the
empirical black hole/bulge correlations.  Each galaxy was modeled as a
Hernquist profile \citep{H90} with a bulge virial mass \citep{M03}
$M_{vir}\approx 3R_{e}\sigma^{2}/G$ chosen according to a log-normal
distribution with a mean given by the relation \citep{M03}
$M_{vir}\approx(1.71 \times 10^{11} M_{\odot}) [M/
(4\times10^{8}~M_{\odot})]^{1.04}$ and standard deviation of 0.25
dex \citep{M03}. The velocity dispersion was calculated from the
M--$\sigma$ relation \citep{T02}, $\sigma$ $\approx$ (262 km s$^{-1}$)
$[M/(4\times 10^{8} M_{\odot})]^{0.249}$, and the bulge scale radius set by
$R_{e}=GM_{vir}/3 \sigma^{2}$.  We then added to the stellar bulge a dark 
matter halo with a mass ratio that is an order of magnitude larger than the 
cosmic matter-to-baryon ratio (corresponding to a star formation efficiency 
of 10\%) and a concentration parameter of 9.0($M_{halo}$ / 2.10 $\times$ 
10$^{13}$ $M_{\odot}$)$^{-0.13}$ with a scatter of 0.18 dex \citep{B01}.

The black hole was ejected radially from the galactic center with the final
velocity from the 3-body run and its trajectory integrated for $6 \times
10^{7}$ yr.  We imposed dynamical friction with $\ln\Lambda=3$, which had
almost no effect on the trajectories in high velocity ejections, but
efficiently damped out those with low velocities.  Finally, we chose a 
random line of sight and projected all distances and velocities 
accordingly.  We added up the total time spent in each logarithmic distance 
and velocity bin to assess the relative probability of seeing the quasar at 
different projected separations and line-of-sight velocities.  Figure 2 
shows the resulting probability distributions.  The observed parameters of 
the quasar under consideration, $D\sim 7.2$ kpc and $v \la 200~{\rm km
\hspace{1.5pt} s}^{-1}$, appear to be fairly typical in the statistics of
our runs.  Note that the results scale simply with the black hole mass
in our model.  If the inner binary always starts at a fixed fraction of the
hardening radius, then $M_{bh}\propto \sigma^4$ gives $a_{bin}
\propto M_{bh}/\sigma^{2} \propto M_{bh}^{1/2}$.  We then have $v_{ej}
\propto \sqrt{M_{bh}/a_{bin}} \propto M_{bh}^{1/4} \propto
\sigma$.  The self-similarity of the Hernquist model implies that the
turnaround distance $r_{max}$ in units of the galactic scale radius is
fixed by $v_{ej}$ in units of $\sigma$.  Since $v_{ej}/\sigma$ is constant,
we get $r_{max} \propto R_{e} \propto M_{bh}^{1/2}$.  Hence neglecting
minor deviations from self-similarity due to dynamical friction, we expect
the velocities in our plots to scale roughly as $M_{bh}^{1/4}$ and
distances as $M_{bh}^{1/2}$.

\begin{figure} 
\plotone{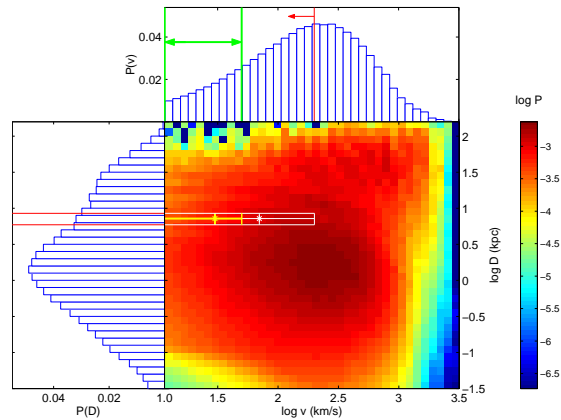} 
\caption{Probability $P$ of observing the black hole at various projected
separations $D$ and line-of-sight velocities $v$.  The colors are scaled in
$\log P(D,v)$, where $P(D,v)$ is the probability density for obtaining the
values $(D,v)$.  The probability incorporates the length of time spent by
the ejected black hole at these values, averaged over $\sim 10^4$
successful runs (which were neither terminated before the ejection nor
eliminated by our minimum-distance criterion). The white star shows the
estimated coordinates of the HE0450-2958 quasar based on the radial
velocity range reported in Magain \etal, with errors bracketed by the white
lines.  The yellow (thicker) lines indicate the magnitude of the quasar's
redshift relative to the galaxy (30$\pm$20 km/s) measured by Merritt \etal
(2005).  The histograms show the corresponding projections of $P$ onto the
$D$ and $v$ axes.}
\end{figure}

\section{Discussion and conclusions}
We have proposed that the HE0450-2958 quasar, around which no galactic
host has so far been detected, was ejected from the nearby galaxy in a
three-body interaction following a merger of a massive gas-rich galaxy with
a giant elliptical.  This scenario explains (a) the fueling of the bright
quasar displaced by $\sim$7 kpc from the galactic center; (b) the survival
of the binary in the gas-poor galaxy before the merger; and (c) the disturbed
appearance of the companion galaxy.  One may also ask how the presence of
the ``blob'' and narrow line emission in the quasar spectrum, both
indicative of ionized gas in the quasar's vicinity far outside the black
hole radius of influence, fit into the ejection picture.  Merritt \etal
(2005) infer a size of roughly 1.5 kpc for the narrow-line region (NLR).
Since the gas at this distance cannot possibly be
bound to the black hole, they argue that the very small observed redshifts
between quasar, galaxy, and narrow emission line gas rule out the ejection 
scenario.  However since the ejected quasar would spend most of its time 
near turnaround, we find a quite small velocity difference between the 
quasar and ULIRG to be consistent with the high-velocity ejection 
model.  For instance, the black hole modelled in Figure 1, with 
$v_{ej} \sim 1200~{\rm km~s^{-1}}$, spends 17\% of its time with $v <$ 
100 km/s. After projecting onto the line of sight and averaging over many 
runs as in Figure 2, one can see that very small velocities are still 
prevalent in our statistics at distances around 7 kpc.

In galaxy merger simulations including feedback from star formation and
AGN \citep{SDH05}, large amounts of gas are blown from the galactic center out
to distances of $\sim$5--10 kpc during the coalescence of the two nuclei.
The disturbed system also shines as a ULIRG during most of this phase.
Once the feedback shuts off, the gas that does not escape settles back into 
the inner galaxy on the dynamical timescale of $\sim$10$^{8}$ yrs.  The 
mass in narrow line emitting gas inferred from absolute line fluxes in 
typical NLR is around 10$^{5-6}M_{\odot}$ \citep{PET97}.  Ten million years 
after the core merger, more than this mass in gas below 10$^{5}$ K would 
likely be available within the few kpc sphere surrounding a black hole 
ejected to a distance of $\sim$7 kpc.  It is plausible that clouds formed 
from this cool gas could produce the observed narrow line emission in the 
quasar spectrum while the black hole (and gas) are near turnaround.  The 
observed blob would be a portion of the gas photoionized by anisotropic 
radiation from the quasar.  It is not feasible for Bondi accretion from this 
external gas to fuel the quasar, as the densities required are 
unrealistically high.  Rather the quasar must be fueled by the remnant inner 
accretion disk dragged with it from the galactic center, which can remain 
bound to it through the ejection as demonstrated above.

Our proposed scenario is viable as long as the merger process does not
induce coalescence of the binary through dynamical friction on a rearranged
distribution of stars and gas, {\it before} the three black holes
interact. For typical parameters, we find that the feeding of the loss cone
of stars by tidal interactions during the merger is not sufficient to cause
coalescence of the binary before the third black hole sinks in \citep{R81}.
Coalescence through friction on gas would be possible by momentum
conservation only if a large quantity of gas, comparable to or larger than
the binary mass, is channeled through the interior of the binary before the
three black holes interact. However, since the gas initially resides within
the quasar's original host galaxy, it is more likely to fuel the quasar
before reaching the nucleus of the dry elliptical that hosts the binary.
Feedback from the quasar could then easily supply sufficient energy to 
prevent the gas which gets tidally-stripped from the outer envelope of its
host galaxy \citep{WL03} from settling into the few parsec scale region 
surrounding the binary before the three black holes interact.

Both our calculations and the identification of HE0450-2958 as a quasar 
without a typical host \citep{M05} are based on the correlations between 
black hole masses and their host bulges established for isolated, quiescent 
galaxies.  The extension of these correlations to active and merging 
systems is somewhat less certain.  Several authors \citep{GH05P, O04, W02} 
have studied the M$_{BH}$--$\sigma$ and M$_{BH}$--M$_{bulge}$ relations in 
active galaxies and reported statistically significant deviations from the 
slope and zero-point measured for inactive systems, as well as slightly 
larger scatter.  However the reported deviations are not typically greater 
than a factor of 2-3 and hence qualitatively support our application of the 
correlations to an interacting system.  

Since the mass of the HE0450-2958 black hole and corresponding expected 
host properties are uncertain, one cannot prematurely conclude that it is
truly a ``naked'' quasar.  Further observations, particularly in bands 
admitting a higher ratio of galaxy to quasar luminosity, are needed to place 
tighter constraints on the properties of any host galaxy.  This paper 
demonstrates the plausibility of a three-body ejection producing a naked 
quasar near a recently-merged galaxy.  If such a system is unambiguously 
observed, this would provide a key link in our understanding of galaxy 
formation and the hierarchical buildup of supermassive black holes.

\acknowledgements

We are grateful to Sverre Aarseth for use of his Chain code and for
enlightening discussions about black hole binaries.  We also thank Thomas
J. Cox, Volker Springel, and Lars Hernquist for providing simulation data
and for useful discussions about the coevolution of black holes and
galaxies.  This work was supported in part by NASA grants NAG 5-13292,
NNG05GH54G, and NSF grant AST-0204514.

\end{document}